\documentclass[11pt,draftcls,onecolumn]{IEEETran}
\usepackage{amssymb, amsmath, graphicx, cite, epsfig, booktabs}

\usepackage[usenames,dvipsnames]{pstricks}
\usepackage{epsfig}
\usepackage{pst-grad} 
\usepackage{pst-plot} 
\usepackage{bm}
\usepackage{mathtools}
\usepackage{tensor}

\usepackage{caption}
\newcommand{\thmend}{\hspace*{\fill}~\IEEEQEDopen\par\endtrivlist\unskip}

\renewcommand{\footnoterule}{%
  \kern -3pt
  \hrule width 150pt height .2pt
  \kern 2pt
}

\newcommand{\bx}[1]{x_{#1}^{n}}

\newcommand{\Bt}{{{\bm{\theta}}}}

\newcommand{\Uo}{{U}^{n}_{1}}
\newcommand{\Ut}{{{{U_{2}^{n}}}}}
\newcommand{\Us}{{U}^{n}_{\mathcal{S}}}
\newcommand{\Usc}{{{{U_{{\mathcal{S}^{c}}}^{n}}}}}
\newcommand{\uo}{{u}^{n}_{1}}

\newcommand{\X}[2]{{{{X}}_{#1}^{#2}}}
\newcommand{\x}[2]{{{{x}}_{#1}^{#2}}}

\newcommand{\Y}[2]{{Y}^{#2}_{#1}}

\newcommand{\dm}{\mathsf{d_{max}}}
\newcommand{\al}{\mathsf{\alpha}}
\newcommand{\ep}{\epsilon}
\newcommand{\de}{\delta}
\newcommand{\mS}{\mathcal{S}}
\newcommand{\mC}{\mathcal{C}}

\newcommand{\mA}{\mathcal{A}}
\newcommand{\mB}{\mathcal{B}}
\newcommand{\mSc}{\mathcal{S}^{c}}

\newtheorem{theorem}{\bf Theorem}
\newtheorem{lemma}[theorem]{\bf Lemma}

\newtheorem{definition}[theorem]{\bf Definition}
\newtheorem{corollary}[theorem]{\bf Corollary}

\newtheorem{remark}[theorem]{\bf Remark}

\newcommand{\qed}{\nobreak \ifvmode \relax \else
      \ifdim\lastskip<1.5em \hskip-\lastskip
      \hskip1.5em plus0em minus0.5em \fi \nobreak
      \vrule height0.35em width0.4em depth0.15em\fi}

\allowdisplaybreaks[4]
\renewcommand{\baselinestretch}{1.49}

\begin{document}


\title{\vspace{-0.5cm} Time-Asynchronous Gaussian Multiple Access Relay Channel with Correlated Sources
\author{{H.~Ebrahimzadeh Saffar, M.~Badiei Khuzani, and P.~Mitran}
\thanks{
The authors are with the Department of Electrical and Computer Engineering, University of Waterloo, Ontario, Canada. Email:~\{h4ebrahi, mbadiei, pmitran\}@uwaterloo.ca.
}
\thanks{
This paper was presented in part at the 2013 IEEE International Symposium on Information Theory (ISIT), Istanbul, Turkey \cite{Saffar_Mitran_ISIT2013}.}} }

\maketitle \thispagestyle{plain}
\pagestyle{plain} 



\maketitle \thispagestyle{plain}
\pagestyle{plain} 
\vspace{-.15cm}
\begin{abstract}

We study the transmission of a set of correlated sources $(U_1,\cdots,U_K)$ over a Gaussian multiple access relay channel with time asynchronism between the encoders. We assume that the maximum possible offset $\dm(n)$ between the transmitters grows
without bound as the block length $n \rightarrow \infty$ while the relative ratio ${\dm(n) / n}$ of the
maximum possible offset to the block length asymptotically vanishes. For such a joint
source-channel coding problem, and under specific gain conditions, we derive necessary and sufficient conditions for reliable communications and show that separate source and
channel coding achieves optimal performance. In particular, we first derive a general outer bound on the
source entropy content for all channel gains as our main result. Then, using Slepian-Wolf source coding combined with the
channel coding scheme introduced in \cite{Cover_McEliece:81} on top of block Markov coding, we show that the thus achieved inner bound
matches the outer bound. Consequently, as a corollary, we also address the problem of sending a pair of correlated sources over a two user interference channel in the same context.

\end{abstract}


\section{Introduction} \label{section_intro}

Time synchronization between nodes of a communication network is a common assumption made to
analyze and design such networks. However, in practice, it is very difficult to exactly
synchronize separate nodes either in time or frequency. As an example, in systems with different
transmitters, the transmitters must use their own locally generated clock. However, the
initialization might be different for each clock and the frequencies at the local signal generators
may not be perfectly matched \cite{Hui_Humblet:85}. Indeed, achieving time, phase or frequency synchronization in practical communication systems has been a major engineering issue and still remains an active area of research (see e.g., \cite{Wornell_asynchronism:2009}). Thus, fundamental limits of communication in the presence of time asynchronism should be explicitly addressed as a tool to better understand and tackle
real-world challenges in the context of multiuser information theory.

The problem of finding the capacity region of multiuser channels with no time synchronization
between the encoders is considered in \cite{Cover_McEliece:81}, \cite{Hui_Humblet:85},
\cite{Farkas_Koi:2012}, and \cite{Grant_Rimoldi_Urbanke_Whiting:2001} from a channel coding
perspective only for the specific case of multiple access channels (MAC). In \cite{Verdu_memory:1989}, a frame asynchronous MAC with memory is considered and it is shown that the capacity region can be drastically reduced in the presence of frame asynchronism.
In \cite{Verdu:1989}, an asynchronous MAC is also considered, but with symbol asynchronism. All of
these works constrain themselves to the study of channel coding only and disregard the source-channel communication
of correlated sources over asynchronous channels. In this paper, we are interested in the problem of joint source-channel coding (JSCC) of a set of correlated sources over time-asynchronous multiuser channels which can include relaying as well. In particular, we focus on the analysis of JSCC for a MAC with the presence of a relay, also known as a multiple access relay channel (MARC).

The problem of JSCC for multiuser networks is open in general. However, numerous results have been published on different aspects of the problem for specific channels and under specific assumptions such as phase or time asynchronism between the nodes. In \cite{Cover_ElGamal_Salehi:1980}, a sufficient condition for lossless communication of correlated sources over a discrete memoryless MAC is given. Although not always optimal, as shown in \cite{Dueck:1981}, the achievable scheme of \cite{Cover_ElGamal_Salehi:1980} outperforms separate source-channel coding. In \cite{FadiAbdallah_Caire:2008}, however, the authors show that under phase fading, separation is optimal for the important case of a Gaussian MAC. Also, \cite{Saffar:phase}, \cite{Saffar_Globecom:2012} show the optimality of separate source-channel coding for several Gaussian networks with phase uncertainty among the nodes. Other authors have derived JSCC coding results for the broadcast channels \cite{Coleman:2006}, \cite{Tian_Diggavi_Shamai_BCfull:2011}, interference relay channels \cite{Saffar_ISIT:2012}, and other multiuser channels \cite{Gunduz:2009}. Furthermore, for lossy source-channel coding, a separation approach is shown in \cite{Tian_Diggavi_Shamai_full_version:2012} to be optimal or approximately optimal for certain classes of sources and networks.

In \cite{Saffar_Mitran_ISIT2013}, we have considered a two user time asynchronous Gaussian MAC with a pair of correlated sources. There, we have derived necessary and sufficient conditions for reliable communication and consequently derived a separation theorem for the problem. This paper extends the work of \cite{Saffar_Mitran_ISIT2013} to a more general setup with $K$ nodes and a relay. Also, the recent work \cite{Yemini_Asynchronous_Side:2014} considers the point-to-point state-dependent and cognitive multiple access channels with time asynchronous side information.

In \cite{Cover_McEliece:81}, the authors have considered a MAC with no common time base between encoders.
There, the encoders transmit with an unknown offset with respect to each other, and the offset is
bounded by a maximum value $\dm(n)$ that is a function of coding block length $n$. Using a
time-sharing argument, it is shown that the capacity region is the same as the capacity of the
ordinary MAC as long as $\dm(n)/n \rightarrow 0$. On the other hand, \cite{Hui_Humblet:85}
considers a {\em totally asynchronous} MAC in which the coding blocks of different users can
potentially have no overlap at all, and thus potentially have several block lengths of shifts between
themselves (denoted by random variables $\Delta_i$). Moreover, the encoders have different clocks
that are referenced with respect to a standard clock, and the offsets between the start of code
blocks for the standard clock and the clock at transmitter $i$ are denoted by random variables
$D_i$. For such a scenario, in \cite{Hui_Humblet:85}, it is shown that the capacity region differs
from that of the synchronous MAC only by the lack of the convex hull operation. In
\cite{Poltyrev:83}, Poltyrev also considers a model with arbitrary delays, known to the receiver
(as opposed to \cite{Hui_Humblet:85}). Among other related works is the recent paper
\cite{Farkas_Koi:2012} that finds a single letter capacity region for the case of a $3$ sender MAC,
$2$ of which are synchronized with each other and both asynchronous with respect to the third one.


In this paper, we study the communication of $K$ correlated sources over a $K$-user Gaussian time-asynchronous MARC
(TA-MARC) where the encoders cannot synchronize the starting times of their codewords. Rather, they
transmit with unknown positive time delays $d_1,d_2,\cdots,d_{K+1}\geq 0$ with respect to a time reference, where the index $K+1$ indicates the relay transmitter. The time shifts are also bounded by $d_{\ell} \leq \dm(n),$ $\ell=1,\cdots,K+1$, where $n$ is the codeword block length. Moreover, we assume that the offsets $d_1,d_2,\cdots,d_{K+1}$ are unknown to the transmitters as a practical assumption since they are not controlled by the transmitters. We further assume that the
maximum possible offset
$\dm(n) \rightarrow \infty$ as $n\rightarrow \infty$ while ${\dm(n) / n} \rightarrow 0$.

The rest of this paper is organized as follows. In Section \ref{section_preliminaries}, we present
the problem statement and preliminaries along with a key lemma that is useful in the derivation of
the converse. In Section \ref{section_converse}, as our main result, the converse part of the capacity theorem (i.e., a theorem stating coinciding necessary and sufficient conditions for reliable source-channel communication) is proved. Then, under specific gain conditions, using separate source and channel coding and the results of \cite{Cover_McEliece:81} combined with block Markov coding, it is shown in Section
\ref{section_achievability} that the thus achievable region matches the outer bound. Section \ref{results_statements} then states a separation theorem under specific gain conditions for the TA-MARC as the combination of converse and achievability parts along with a corollary that results for the interference channel. Finally, Section \ref{section_conclusion} concludes the paper. \vspace{-.2cm}


\section{Problem Statement and a Key Lemma} \label{section_preliminaries}

{\em Notation}: In what follows, we denote random variables by upper case letters, e.g., $X$, their realizations by lower case
letters, e.g., $x$, and their alphabet by calligraphic letters, e.g., $\mathcal{X}$. For integers $0 \leq a \leq b$, $Y_{a}^{b}$ denotes
the $b-a+1$-tuple $(Y[a],\cdots,Y[b])$, and $Y^{b}$ is a shorthand for $\Y{0}{b-1}$. Without confusion, $X_{\ell}^{n}$ denotes the
length-$n$ MARC input codeword $(X_{\ell}[0],\cdots,X_{\ell}[n-1])$ of the $\ell$th transmitter, and based on
this, we also denote $(X_{\ell}[a],\cdots,X_{\ell}[b])$ by $X_{\ell,a}^{b}$. The $n$-length
discrete Fourier transforms (DFT) of the $n$-length codeword $X_{\ell}^{n}$ is denoted by $\hat{X}_{\ell}^{n} =
{\rm{DFT}}(\X{\ell}{n})$. Furthermore, let $[1,K] \triangleq \{1,\cdots,K\}$, for $\forall K \in \mathbb{N}$.

\begin{figure}
\centering
{\includegraphics[keepaspectratio=true, width=12.68cm]{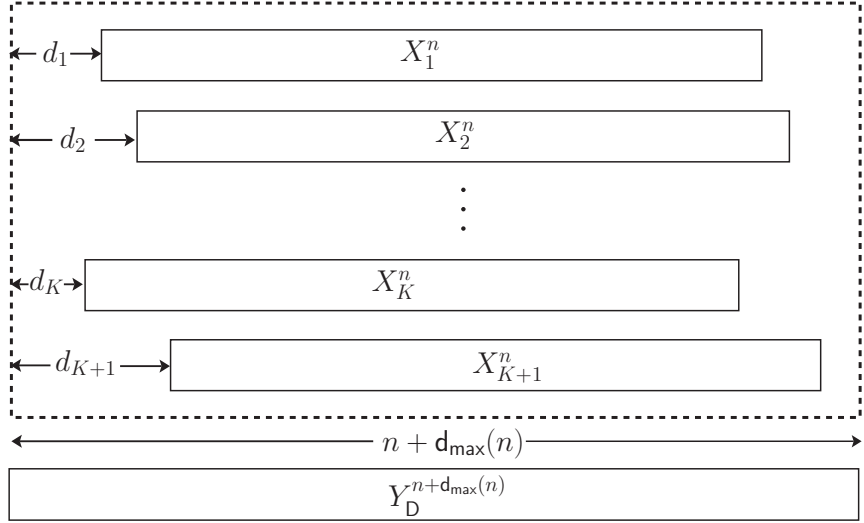}}
\caption{{Gaussian time asynchronous multiple access relay channel (TA-MARC), with delays} $d_{1},\cdots,d_{K+1}$.}
\label{fig:TA-MAC}
\end{figure}

Consider $K$ finite alphabet sources $\{(U_{1}[i],U_{2}[i],\cdots,U_{K}[i])\}_{i=0}^{\infty}$ as correlated random variables drawn according to a distribution $p(u_1,u_2,\cdots,u_K)$. The sources are memoryless, i.e.,
$(U_{1}[i],U_{2}[i],\cdots,U_{K}[i])$'s are independent and identically distributed (i.i.d) for $i=1,2,\cdots$. The indices $1,\cdots,K$, represent the transmitter nodes and the index $K+1$ represents the relay transmitter. All of the sources are to
be transmitted to a destination by the help of a relay through a continuous alphabet, discrete-time memoryless
multiple-access relay channel (MARC)
with time asynchronism between different transmitters and the relay. Specifically, as depicted in Fig. \ref{fig:TA-MAC}, the encoders use different time references and thus we assume that the encoders start transmitting with offsets of
\begin{align}
0 \leq d_{\ell} \leq \dm(n), \quad \ell=1,\cdots,K+1,
\end{align}
symbols with respect to a fixed time reference, where $d_{K+1}$ is the offset for the relay transmitter with respect to the time reference. 

Hence, the probabilistic characterization of the time-asynchronous Gaussian MARC, referred to as a Gaussian TA-MARC and denoted by $\mathcal{M}([1,K+1])$ throughout the paper, is
described by the relationships
\begin{align}\label{channel-model-1}
Y_{\mathsf{D}}[i] & = \sum_{\ell=1}^{K+1} g_{\ell\mathsf{D}}X_{\ell}[i-d_{\ell}] + Z_{\mathsf{D}}[i], \quad i=0,1,\cdots,n+\dm(n)-1,
\end{align}
\noindent as the $i$th entry of the received vector $Y_{\sf{D}}^{n+\dm(n)}$ at the destination ($\sf{D}$), and
\begin{align}\label{channel-model-2}
Y_{\mathsf{R}}[i] & = \sum_{\ell=1}^{K} g_{\ell \mathsf{R}}X_{\ell}[i-d_{\ell}] + Z_{\mathsf{R}}[i], \quad i=0,1,\cdots,n+\dm(n) - 1,
\end{align}
\noindent as the $i$th entry of the received vector $Y_{\mathsf{R}}^{n+\dm(n)}$ at the relay ($\mathsf{R}$), where
\begin{itemize}
\item $g_{\ell \mathsf{D}},\ell=1,\cdots,K+1,$ are complex gains from transmission nodes as well as the relay (when $\ell=K+1$) to the destination, and $g_{\ell\mathsf{R}}, \ell = 1,\cdots,K,$ are complex gains from the transmission nodes to the relay,
\item $X_{\ell}[i-d_{\ell}], \ell=1,\cdots,K+1$, are the delayed channel inputs  such that $X_{\ell}[i-d_{\ell}] = 0$ if $(i-d_{\ell})$$ \notin \{0,1,\cdots,n-1\}$ and $X_{\ell}[i-d_{\ell}]\in \mathbb{C}$ otherwise,
\item $Z_{\mathsf{D}}[i],Z_{\mathsf{R}}[i] \sim {\mathcal{C}\mathcal{N}(0,N)}$ are circularly symmetric complex Gaussian noises at the destination and relay, respectively.
\end{itemize}
Fig. \ref{fig:TA-MAC} depicts the delayed codewords of the encoders, and the formation of the received codeword for the TA-MARC.



We now define a joint source-channel code and the notion of reliable communication for a Gaussian
TA-MARC in the sequel.

\begin{definition}
A block joint source-channel code of length $n$ for the Gaussian TA-MARC with the block of correlated
source outputs $$\{(U_1[i],U_2[i],\cdots,U_K[i])\}_{i=0}^{n-1}$$ is defined by
\begin{enumerate}
\item {A set of encoding functions with the bandwidth mismatch factor of unity\footnote{The assumption of unity mismatch factor is without loss of generality and for simplicity of exposition. Extension to the more general setting with different mismatch factors can be achieved by a simple modification (cf. Remark \ref{Remark:about_mismatch}).}, i.e.,
\begin{align*}
f_{\ell}^{n}&: \mathcal{U}_{\ell}^n \rightarrow \mathbb{C}^n, \quad \ell=1,2,\cdots,K,
\end{align*}
\noindent that map the source outputs to the codewords, and the relay encoding function
\begin{align}\label{Eq:relay_encoding_function}
x_{(K+1)}^{i+1} = f_{(K+1)}^{i+1}(y_{\mathsf{R}}[0],y_{\mathsf{R}}[1],\cdots,y_{\mathsf{R}}[i]), \quad i=0,2,\cdots,n-2.
\end{align}
\noindent The sets of encoding functions are denoted by the {\em codebook} $\mathcal{C}^{n} = \Big\{f_{1}^{n},\cdots,f_{K}^{n},\{f_{(K+1)}^{i+1}\}_{i=0}^{n-2}\Big\}$}.

\item{Power constraints $P_\ell$, $\ell=1,\cdots,K+1,$ on the codeword vectors $X^{n}_{\ell}$, i.e.,
\begin{align}\label{Power_constraint}
\mathbb{E}\left[{1 \over n} \sum_{i=0}^{n-1}\vert X_{\ell }[i]\vert^2\right] =
\mathbb{E}\left[{1 \over n} \sum_{i=0}^{n-1}\vert \hat{X}_{\ell }[i]\vert^2\right] \leq P_\ell, \ \
\end{align}
\noindent for $\ell=1,\cdots,K+1$ where we recall that $\hat{X}^{n}_{\ell}=\text{DFT}\{X_{\ell}^{n}\}$, and $\mathbb{E}[\cdot]$ represents the expectation operator.} 
\item{A decoding function $g^n(y_{\sf{D}}^{n+\dm} \vert d_{1}^{K+1}) : \mathbb{C}^{n+\dm} \times [0,\dm]^{K+1} \rightarrow  \mathcal{U}_{1}^n \times\cdots \times \mathcal{U}_{K}^n. $ }
\end{enumerate}
\end{definition}
\begin{definition} \label{reliability_definition} We say the source $\{(U_{1}[i],U_{2}[i],\cdots,U_{K}[i])\}_{i=0}^{n-1}$ of i.i.d. discrete random variables with joint probability mass function $p(u_1,u_2,\cdots,u_K)$ {\em can be reliably sent} over a Gaussian TA-MARC, if there exists a sequence of codebooks $\mathcal{C}^{n}$ and decoders $g^n$ in $n$ such that the output sequences $\Uo,\Ut,\cdots,U_{K}^{n}$ of the source can be estimated from $Y_{\mathsf{D}}^{n+\dm(n)}$ with arbitrarily asymptotically small probability of error uniformly over {\em all} choices of delays $0 \leq $$ d_{\ell} $$\leq \dm(n),$ $\ell=1,\cdots,K+1$, i.e.,
\begin{align}\label{main_error_probability}
\sup_{0 \leq d_{1}, \cdots, d_{K+1} \leq \dm(n)} P_e^n(d_{1}^{K+1}) \longrightarrow 0, \ \ {\rm as}   \  \ n \rightarrow \infty,
\end{align}
\noindent where
\begin{align}
P_e^n(d_{1}^{K+1}) \triangleq P[g(Y_{\sf{D}}^{n+\dm(n)} \vert d_{1}^{K+1}) \neq (\Uo,\Ut,\cdots,U_{K}^{n}) \vert d_{1}^{K+1}],
\end{align}
is the error probability for a given set of offsets $d_{1}^{K+1}$. \thmend
\end{definition}

We now present a key lemma that plays an important role in the derivation of our results. In order
to state the lemma, we first need to define the notions of a {\em sliced} MARC and a {\em sliced cyclic} MARC as follows:
%

\begin{definition}
Let $\mS \subseteq [1,K+1]$ be a subset of transmitter node indices. A Gaussian sliced TA-MARC ${\mathcal{M}}(\mS)$ corresponding to the Gaussian TA-MARC ${\mathcal{M}}([1,K+1])$ defined by \eqref{channel-model-1}-\eqref{channel-model-2}, is a MARC in which only the codewords of the encoders with indices in $\mS$ contribute to the destination's received signal, while the received signal at the relay is the same as that of the original Gaussian TA-MARC $\mathcal{M}([1,K+1])$.

In particular, for the Gaussian sliced MARC ${\mathcal{M}}(\mS)$, the received signals at the destination and the relay at the $i$th time index, denoted by ${Y}_{\mathsf{D}(\mS)}[i]$ and ${Y}_{\mathsf{R}(\mS)}[i]$ respectively, are given by
\begin{align}\label{sliced_destination}
{Y}_{\mathsf{D}(\mS)}[i] = \sum_{\ell \in \mS} g_{\ell\mathsf{D}}X_{{{\ell}}}[i-d_{\ell}] + {Z}_{\mathsf{D}}[i], \quad {i=0,\cdots,n+\dm-1},
\end{align}
\noindent and
\begin{align}\label{sliced_relay}
{Y}_{\mathsf{R}(\mS)}[i] = Y_{\mathsf{R}}[i], \quad {i=0,\cdots,n+\dm-1}.
\end{align}
\end{definition}

\begin{figure}
\hspace{-1.5cm}\centering{\includegraphics[keepaspectratio = true, height=14.2cm]{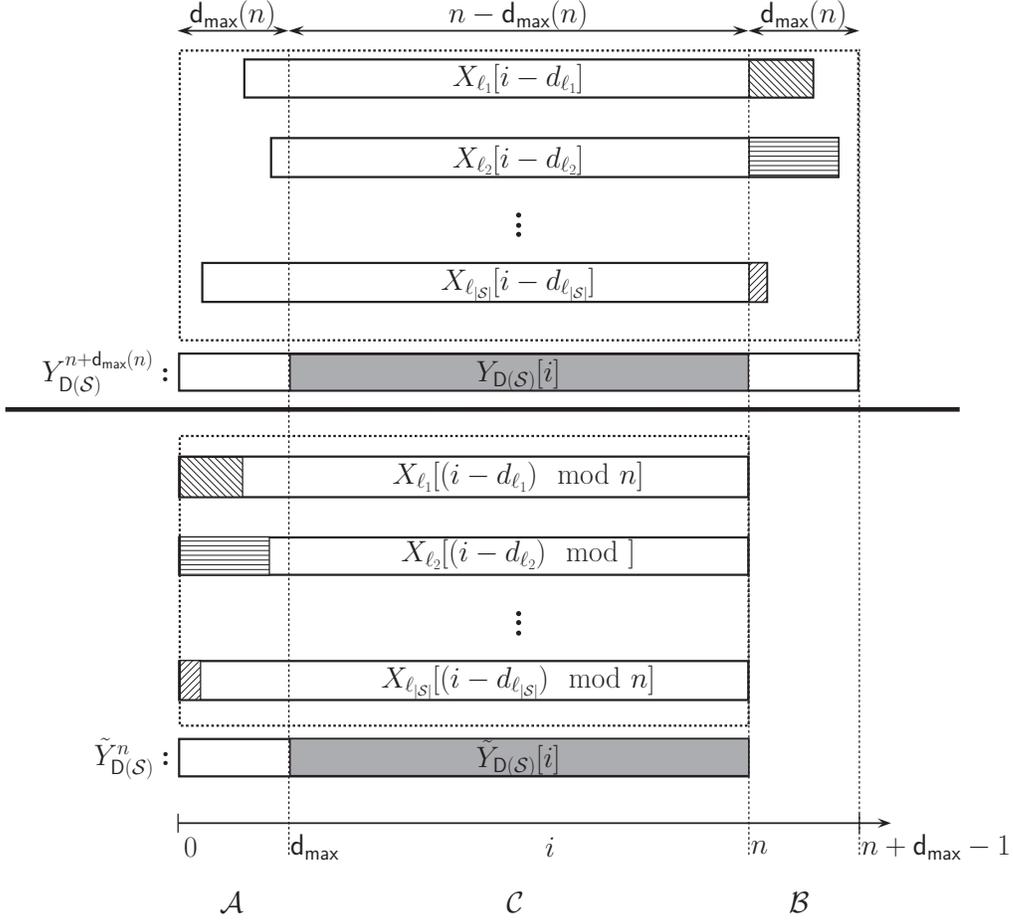}}
\caption{Codewords of a Gaussian sliced TA-MARC $\mathcal{M}(\mS)$ (top) and the corresponding sliced cyclic MARC $\tilde{\mathcal{M}}(\mS)$ (bottom).}
\label{fig:partial-TA-MAC}
\end{figure}

\begin{definition}
A sliced cyclic MARC $\widetilde{\mathcal{M}}(\mS)$, corresponding to the sliced TA-MARC $\mathcal{M}(\mS)$ defined by \eqref{sliced_destination}-\eqref{sliced_relay}, is a sliced TA-MARC in which the codewords are cyclicly shifted around the $n$th time index to form new received signals at the destination \textit{only}. Specifically, the corresponding outputs of the sliced cyclic MARC $\widetilde{\mathcal{M}}(\mS)$ at the destination and the relay at the $i$th time index, denoted by $\tilde{Y}_{\mathsf{D}(\mS)}[i]$ and $\tilde{Y}_{\mathsf{R}(\mS)}[i]$ respectively, can be written as
\begin{align}
\tilde{Y}_{\mathsf{D}(\mS)}[i] = \sum_{\ell \in \mS} g_{\ell \mathsf{D}}X_{{\ell}}[(i-d_{\ell})\hspace{0mm}\mod n] + Z_{\mathsf{D}}[i], \quad i = 0,\cdots,n-1,
\end{align}
\noindent and
\begin{align}
\tilde{Y}_{\mathsf{R}(\mS)}[i] &= \sum_{\ell = 1}^{K} g_{\ell \mathsf{R}}X_{{{\ell}}}[i-d_{\ell}] + Z_{\mathsf{R}}[i], \quad i = 0,\cdots,n-1,
\nonumber\\
&=Y_{\mathsf{R}}[i].
\end{align}

In particular, as shown in Fig. \ref{fig:partial-TA-MAC}, the tail of the codewords are cyclicly shifted to the beginning of the block, where the start point of the block is aligned with the first time instant. The destination's output $\tilde{Y}_{\mathsf{D}(\mS)}^{n}$ of the sliced cyclic MARC is the $n$-tuple that results by adding the shifted versions of the codewords $X_{\ell}^{n},{\ell \in \mS}$. As indicated in Fig. \ref{fig:partial-TA-MAC}, we divide the entire time interval $[0,n+\dm-1]$ into three subintervals $\mA, \mB$, and $\mC$ where
\begin{itemize}
\item $\mA$ is the sub-interval representing the left tail of the received codeword, {i.e.}, $[0,\dm-1]$,
\item $\mB$ represents the right tail, {i.e.}, $[n,n+\dm-1]$,
\item $\mC$ represents a common part between the sliced TA-MARC and sliced cyclic MARC, {i.e.}, $[\dm,n-1]$.
\end{itemize}
\end{definition}

\begin{remark}
\label{Remark_01}
In both sliced TA-MARC and sliced cyclic MARC, the observation $Y^{n+\dm}_{\mathsf{R}}$ of the relay remains unchanged. Therefore, the generated channel input at the relay $X_{K+1}^{n}$ is the same as the original TA-MARC due to \eqref{Eq:relay_encoding_function} when the same relay encoding functions are used.
\end{remark}

The following lemma implies that, for every choice of $\mS \subseteq [1,K+1]$, the mutual information rate between the inputs and the destination's output in the Gaussian sliced TA-MARC $\mathcal{M}(\mS)$ and the sliced cyclic MARC $\widetilde{\mathcal{M}}(\mS)$ are asymptotically the same, i.e., their difference asymptotically vanishes. This fact will be useful in the analysis of the problem in Section
\ref{section_converse}, where we can replace a sliced TA-MARC with the corresponding sliced cyclic MARC.

Before stating and proving the key lemma, we define the following notations:
\begin{align}
Y_{\mathsf{D}{(\mS)}}[\mA] & \triangleq \{Y_{\mathsf{D}{(\mS)}}[i]: i \in \mA \}, \\
\tilde{Y}_{\mathsf{D}{(\mS)}}[\mA]& \triangleq \{\tilde{Y}_{\mathsf{D}{(\mS)}}[i]: i \in \mA \}, \\ \label{Eq:definitionofXs}
X_{\mathcal{S}}^{n} & \triangleq \{X_{\ell}^{n}: \ell \in \mS\}, \\
\vec{X}_{\mathcal{S}}[{\mA}] & \triangleq \{X_{\ell}[i-d_{\ell}]: \ell\in \mS, i \in {\mA}\},\\
\tilde{\vec{X}}_{\mathcal{S}}[{\mA}] & \triangleq \{X_{\ell}[i-d_{\ell}\ \text{mod}\ n]: \ell\in \mS, i \in {\mA}\},
\end{align}
\noindent where $\mS \subseteq [1,K+1]$ is an arbitrary subset of transmitter nodes indices, and recall that $X_{\ell}[i-d_{\ell}] = 0$, for $i-d_{\ell} \not\in \{0,1,\cdots,n-1\}$. Similarly, we can define $Y_{\mathsf{D}{(\mS)}}[\mB]$, $Y_{\mathsf{D}{(\mS)}}[\mC]$, $\tilde{Y}_{\mathsf{D}{(\mS)}}[\mB],\cdots$, by replacing $\mA$ with $\mB$ or $\mC$ in the above definitions.

\begin{lemma} \label{Key_lemma} For a Gaussian sliced TA-MARC $\mathcal{M}(\mS)$, and the corresponding sliced cyclic MARC $\widetilde{\mathcal{M}}(\mS)$,
\begin{align}
{1 \over n} \left| I(\X{\mS}{n};{Y}_{\mathsf{D}(\mS)}^{n+\dm} \vert {d_{1}^{K+1}}) -
I(\X{\mS}{n}; \tilde{Y}^{n}_{\mathsf{D}(\mS)}  \vert {d_{1}^{K+1}}) \right| & \leq \ep_{n}, \quad \forall \ d^{K+1}_{1} \in [0,\dm(n)]^{K+1},  \label{lemma_main_expression}
\end{align}
\noindent for all $\mS \subseteq [1,K+1]$, where $\ep_{n}$ does not depend on $d_{1}^{K+1}$ and $\ep_{n} \rightarrow
0$, as $n\rightarrow \infty$. \thmend  
\end{lemma}
\begin{IEEEproof}

Noting that the mutual information between subsets of two random vectors is a lower bound on the mutual information between the original random vectors, we first lower bound the original mutual information $I(\X{\mS}{n};{Y}_{\mathsf{D}(\mS)}^{n+\dm} \vert {d_{1}^{K+1}})$:
\begin{align}
& I(\vec{X}_{\mS}[\mC]; Y_{{\mathsf{D}(\mS)}}[\mC]\vert {d_{1}^{K+1}}) \leq I(\X{\mS}{n};{Y}_{\mathsf{D}(\mS)}^{n+\dm} \vert {d_{1}^{K+1}}). \label{first_MI_lower}
\end{align}
\noindent Then, by splitting the entropy terms over the intervals $\mA, \mB$, and $\mC$ as depicted in Fig. \ref{fig:partial-TA-MAC}, we upper bound the same mutual information term $I(\X{\mS}{n}$$;{Y}_{\mathsf{D}(\mS)}^{n+\dm}$$ \vert {d_{1}^{K+1}})$ as follows:
\begin{align}
 I(\X{\mS}{n};{Y}_{\mathsf{D}(\mS)}^{n+\dm} \vert {d_{1}^{K+1}}) & = h({Y}_{\mathsf{D}(\mS)}^{n+\dm} \vert {d_{1}^{K+1}}) - h({Y}_{\mathsf{D}(\mS)}^{n+\dm} \vert \X{\mS}{n}, {d_{1}^{K+1}}) \nonumber \\
& \leq h(Y_{{\mathsf{D}(\mS)}}[\mA] \vert {d_{1}^{K+1}}) + h(Y_{{\mathsf{D}(\mS)}}[\mB] \vert {d_{1}^{K+1}}) + h(Y_{{\mathsf{D}(\mS)}}[\mC] \vert {d_{1}^{K+1}}) - \sum_{i=0}^{n+\dm-1} h(Z_{\mathsf{D}}[i]) \nonumber \\
& = I(\vec{X}_{\mS}[\mA];Y_{\mathsf{D}{(\mS)}}[{\mA}] \vert {d_{1}^{K+1}}) + I(\vec{X}_{\mS}[\mB];Y_{\mathsf{D}{(\mS)}}[{\mB}] \vert {d_{1}^{K+1}})
+ I(\vec{X}_{\mS}[\mC];Y_{\mathsf{D}{(\mS)}}[{\mC}] \vert {d_{1}^{K+1}}).  \label{first_MI_upper}
\end{align}

Also, the mutual information term $I(\X{\mS}{n}; \tilde{Y}^{n}_{\mathsf{D}(\mS)}  \vert {d_{1}^{K+1}})$ which is associated to the sliced cyclic MARC can be similarly lower bounded as
\begin{align}
& I(\tilde{\vec{X}}_{\mS}[\mC]; \tilde{Y}_{\mathsf{D}{(\mS)}}[\mC]\vert {d_{1}^{K+1}}) \leq I(\X{\mS}{n}; \tilde{Y}^{n}_{\mathsf{D}(\mS)}  \vert {d_{1}^{K+1}}), \label{second_MI_lower}
\end{align}
\noindent and upper bounded as
\begin{align}
I(X^{n}_{\mS}; \tilde{Y}_{\mathsf{D}{(\mS)}}\vert {d_{1}^{K+1}}) & = h(\tilde{Y}_{\mathsf{D}{(\mS)}} \vert {d_{1}^{K+1}}) - h(\tilde{Y}_{\mathsf{D}{(\mS)}} \vert X^{n}_{\mS}, {d_{1}^{K+1}}) \nonumber \\
& \leq h(\tilde{Y}_{\mathsf{D}{(\mS)}}[\mA] \vert {d_{1}^{K+1}}) + h(\tilde{Y}_{\mathsf{D}{(\mS)}}[\mC] \vert {d_{1}^{K+1}})- {\sum_{i=0}^{n-1}} h(Z_{\mathsf{D}}[i]) \nonumber \\ \nonumber
& = I(\tilde{\vec{X}}_{\mS}[\mA]; \tilde{Y}_{\mathsf{D}{(\mS)}}[\mA]\vert {d_{1}^{K+1}})+ I(\tilde{\vec{X}}_{\mS}[\mC]; \tilde{Y}_{\mathsf{D}{(\mS)}}[\mC]\vert {d_{1}^{K+1}})\\
& = I(\tilde{\vec{X}}_{\mS}[\mA]; \tilde{Y}_{\mathsf{D}{(\mS)}}[\mA]\vert {d_{1}^{K+1}})+ I(\vec{X}_{\mS}[\mC]; Y_{\mathsf{D}{(\mS)}}[\mC]\vert {d_{1}^{K+1}}),
\label{second_MI_upper}
\end{align}
\noindent where in the last step, we used the fact that for any $\mathcal{S}\subseteq [1,K+1]$, ${\tilde{Y}}_{\mathsf{D}{(\mS)}}[{\mC}] = Y_{{\mathsf{D}(\mS)}}[{\mC}]$ and $\tilde{\vec{X}}_{\mS}[\mC]=\vec{X}_{\mS}[\mC]$, as there is no cyclic foldover for $i\in {\mC}$.

Hence, combining \eqref{first_MI_lower}-\eqref{first_MI_upper}, and \eqref{second_MI_lower}-\eqref{second_MI_upper}, we can
now bound the difference between the mutual information terms as
\begin{align}
&{1 \over n} \left| I(\X{\mS}{n};{Y}_{\mathsf{D}(\mS)}^{n+\dm} \vert {d_{1}^{K+1}}) -
I(\X{\mS}{n}; \tilde{Y}^{n}_{\mathsf{D}(\mS)}  \vert {d_{1}^{K+1}}) \right| \nonumber \\
& \quad \leq  {1\over n}I(\vec{X}_{\mS}[\mA]; Y_{\mathsf{D}{(\mS)}}[\mA]\vert {d_{1}^{K+1}}) + {1 \over n} I(\vec{X}_{\mS}[\mB]; Y_{\mathsf{D}{(\mS)}}[\mB]\vert {d_{1}^{K+1}}) +{1 \over n} I(\tilde{\vec{X}}_{\mS}[\mA]; \tilde{Y}_{\mathsf{D}{(\mS)}}[\mA]\vert {d_{1}^{K+1}}).  \label{three_terms_expansion}
\end{align}
\noindent But all of the terms in the right hand side of \eqref{three_terms_expansion} can also be
bounded as follows. Consider the first term:
\begin{align}
\nonumber
{1\over n}I(\vec{X}_{\mS}[\mA]; Y_{\mathsf{D}{(\mS)}}[\mA]\vert {d_{1}^{K+1}})& = {1\over n} \left[ h(Y_{\mathsf{D}{(\mS)}}[\mA] \vert d_{1}^{K+1}) -  h(Z_{\mathsf{D}}[{\mA}]) \right]   \\ \nonumber
& \leq {1\over n} \sum_{i \in \mA} \left[ h(Y_{{\mathsf{D}(\mS)}}[i] \vert d_{1}^{K+1}) -  h(Z_{\mathsf{D}}[i]) \right]  \\ \nonumber
& = {1 \over n} \sum_{i \in \mA}  \left[ h\left(\sum_{\ell \in \mS} g_{{\ell}\mathsf{D}}X_{\ell}[i-d_\ell] + Z_{\mathsf{D}}[i] \right)
-   h(Z_{\mathsf{D}}[i]) \right] \\ \nonumber
& \stackrel{\rm (a)}{\leq} {1 \over n} \sum_{i \in \mA}  \log\left(1 +{ {\mathbb{E}\left\vert \sum_{\ell \in \mS} g_{{\ell}\mathsf{D}}X_{\ell}[i-d_\ell] \right\vert}^{2}  \over {N}   } \right)\\ \nonumber
& \stackrel{\rm (b)}{\leq} {1 \over n} \sum_{i \in \mA}  \log\left(1 +{ {\sum_{\ell \in \mS} \vert g_{\ell\mathsf{D}}\vert ^{2} \cdot \sum_{\ell\in \mathcal{S}}\mathbb{E}\vert X_{\ell}[i-d_\ell]\vert^{2}}  \over {N}   } \right)\\ \nonumber
& \stackrel{\rm (c)} \leq {\vert \mA \vert  \over n} \log\left(1 +{ { \sum_{i \in \mA} \left[ {\sum_{\ell \in \mS} \vert g_{\ell\mathsf{D}}\vert ^{2} \cdot \sum_{\ell\in \mathcal{S}}\mathbb{E}\vert X_{\ell}[i-d_\ell]\vert^{2}} \right] }  \over {\vert \mA \vert N}   } \right)\\ \nonumber
& \stackrel{\rm (d)}{=} {\dm  \over n} \log\left(1 +{ { {\sum_{\ell \in \mS} \vert g_{\ell\mathsf{D}}\vert ^{2} \cdot \sum_{\ell\in \mathcal{S}}\mathbb{E}\left[\sum_{i \in \mA}\vert X_{\ell}[i-d_\ell]\vert^{2}\right]}}  \over {\dm N}   } \right)\\ \nonumber
& \leq {\dm  \over n} \log\left(1 +{ { {\sum_{\ell \in \mS} \vert g_{\ell\mathsf{D}}\vert ^{2} \cdot \sum_{\ell\in \mathcal{S}}\mathbb{E}\sum_{i=0}^{n-1}\vert X_{\ell i}\vert^2}}  \over {\dm N}   } \right)\\ \nonumber
&\stackrel{\rm (e)}{\leq} {\dm  \over n} \log\left(1 +{n\over \dm}{ {\sum_{\ell \in \mS} \vert g_{\ell\mathsf{D}}\vert ^{2} \cdot \sum_{\ell\in \mathcal{S}}P_{\ell}}  \over { N}   } \right) \\
& \triangleq \gamma\left(\dfrac{\dm}{n}\right), \label{lemma_inequality_1}
\end{align}
\noindent where $\rm{(a)}$ follows by the fact that Gaussian distribution maximizes the differential entropy \cite[Thm. 8.4.1]{Cover:2006}, $\rm{(b)}$ follows from the Cauchy-Schwartz inequality:
\begin{align}
\left \vert \sum_{\ell\in \mathcal{S}} g_{\ell\mathsf{D}}{X}_{\ell}[i-d_{\ell}] \right\vert^{2} & \leq {\left(\sum_{\ell\in \mathcal{S}}\vert g_{\ell\mathsf{D}} \vert ^{2} \right)} \left(\sum_{\ell\in \mathcal{S}}  \vert X_{\ell}[i-d_{\ell}] \vert^{2}\right), \label{eq_Cauchy}
\end{align}
\noindent $(\rm c)$ follows from concavity of the $\log$ function, $(\rm d)$ follows from the fact that $\vert \mA \vert = \dm$, and $(\rm e)$ follows from the power constraint in \eqref{Power_constraint}.

Similarly, for the second term in the right hand side of \eqref{three_terms_expansion}, it can be shown that
\begin{align}
{1 \over n} I(\vec{X}_{\mS}[\mB]; Y_{\mathsf{D}{(\mS)}}[\mB]\vert {d_{1}^{K+1}})  \leq \gamma\left({\dm\over n}\right). \label{lemma_inequality_2}
\end{align}

Following similar steps that resulted in \eqref{lemma_inequality_1}, we now upper bound the third term in the right hand side of \eqref{three_terms_expansion} as follows
\begin{align}
\nonumber
{1 \over n} I(\tilde{\vec{X}}_{\mS}[\mA]; \tilde{Y}_{\mathsf{D}{(\mS)}}[\mA]\vert {d_{1}^{K+1}})& = {1\over n} \left[ h(\tilde{Y}_{\mathsf{D}{(\mS)}}[\mA] \vert d_{1}^{K+1}) -  h(Z_{\mathsf{D}}[{\mA}]) \right]   \\ \nonumber
& \leq {1\over n}  \sum_{i \in \mA} \left[ h(\tilde{Y}_{\mathsf{D}}[i] \vert d_{1}^{K+1}) -  h(Z_{\mathsf{D}}[i]) \right]  \\ \nonumber
& = {1 \over n}  \sum_{i \in \mA}  \left[ h\left(\sum_{\ell \in \mS} g_{\ell\mathsf{D}}X_{{{\ell}}}[(i-{d_{{\ell}}})\hspace{-2mm} \mod n] + Z_{\mathsf{D}}[i] \Big\vert d_{1}^{K+1}\right) -  h(Z_{\mathsf{D}}[i]) \right] \\ \nonumber
& \leq {1 \over n} \sum_{i \in \mA}  \log\left(1 +{ {\mathbb{E}\left\vert \sum_{\ell \in \mS} g_{{\ell}\mathsf{D}}X_{\ell}[(i-{d_{{\ell}}})\hspace{-2mm} \mod n] \right\vert}^{2}  \over {N}   } \right)
\\ \nonumber
& \leq  {\dm  \over n} \log\left(1 +{n\over \dm}{ {\sum_{\ell \in \mS} \vert g_{\ell\mathsf{D}}\vert ^{2} \cdot \sum_{\ell\in \mathcal{S}}P_{\ell}}  \over { N}   } \right)\\
& = \gamma\left(\dfrac{\dm}{n}\right). \label{lemma_inequality_3}
\end{align}

Based on \eqref{lemma_inequality_1}, \eqref{lemma_inequality_2}, and \eqref{lemma_inequality_3}, the absolute difference between the mutual informations in \eqref{lemma_main_expression} is upper bounded by $3\gamma(\dm/n)$. 
One can see that $3\gamma\left(\dm(n)/n\right) \rightarrow 0$ as $n \rightarrow \infty$,  since for any $a>0$, $z_{n}\log(1 + a/{z_{n}}) \rightarrow 0$ as
$z_{n} \rightarrow 0$, and the lemma is proved by taking $z_{n}=\dm(n)/n$ and $a={\sum_{\ell \in \mS} \vert g_{\ell\mathsf{D}}\vert ^{2}\sum_{\ell\in \mathcal{S}}P_{\ell}}/N$.
\end{IEEEproof}

\vspace{-.4cm}
\section{Converse}\label{section_converse}

\begin{lemma}
\label{Lemma:Reliable_Communication}
Consider a Gaussian TA-MARC with power constraints
$P_1,P_2,\cdots,P_K$ on the transmitters, and the power constraint $P_{K+1}$ on the relay, and the set of encoders' offsets $d_{1}^{K+1}$. Moreover, assume that the set of offsets $d_{1}^{K+1}$ are known to the receiver, $\dm(n) \rightarrow \infty$, and ${\dm(n)
/ n} \rightarrow 0$ as $n\rightarrow \infty$. Then, a necessary condition for reliably
communicating a source tuple $(U^{n}_1,U^{n}_2,\cdots,U^{n}_{K}) \sim {\prod_{i=0}^{n-1}}p(u_{1}[i],u_{2}[i],\cdots,u_{K}[i])$, over such a Gaussian
TA-MARC, in the sense of Definition \ref{reliability_definition}, is given by
\begin{align}
H(U_{\mS}\vert U_{\mSc}) & \leq \log\left(1+{\sum_{\ell \in \mS}\vert g_{\ell\mathsf{D}}\vert ^{2}P_{\ell} \over N}\right),\quad \forall \mS \subseteq [1,K+1] \label{separation_TAMAC_1}
\end{align}
\noindent where $\mS$ includes the relay, {i.e.}, $\{K+1\}\in \mS$, where by definition $U_{K+1}\triangleq \emptyset$, and $\mathcal{S}^{c}\triangleq[1,K+1]/\{\mathcal{S}\}$.
\thmend
\end{lemma}
\begin{remark}
\label{Remark:about_mismatch}
The result of \eqref{separation_TAMAC_1} can be readily extended to the case of mapping blocks of source outputs of the length $m_{n}$ to channel inputs of the length $n$. In particular, for the bandwidth mismatch factor $\kappa \triangleq \lim_{n \rightarrow \infty} {n \over {m_{n}}}$, the converse result in \eqref{separation_TAMAC_1}, to be proved as an achievability result in Section \ref{section_achievability} as well, can be generalized to
\begin{align}
H(U_{\mS}\vert U_{\mSc}) & \leq \kappa \log\left(1+{\sum_{\ell \in \mS}\vert g_{\ell\mathsf{D}}\vert ^{2}P_{\ell} \over N}\right), \quad \forall \mS \subseteq [1,K+1].
\end{align}
Since considering a general mismatch factor $\kappa>0$ obscures the proof, in the following, without essential loss of generality, we present the proof for the case of $\kappa=1$.
\end{remark}
\begin{IEEEproof}

First, fix a TA-MARC with given offset vector $d_{1}^{K+1}$, a codebook $\mathcal{C}^{n}$, and
induced {\em empirical} distribution
\[p(\uo,\cdots,u_{K}^{n},\x{1}{n},\cdots,\x{K+1}{n},y_{\mathsf{R}}^{n+\dm},y_{\mathsf{D}}^{n+\dm} \vert d_{1}^{K+1}).\]
Since for this fixed choice of the offset vector $d_{1}^{K+1}$, $P^n_e(d_{1}^{K+1}) \rightarrow 0$, from Fano's inequality, we have
\begin{align}
{1 \over n}H(\Uo,\Ut,\cdots,U_{K}^{n} \vert Y_{\sf{D}}^{n+\dm},d_{1}^{K+1}) \leq {1 \over n}{P_e^n(d_{1}^{K+1})} \log \|{{\mathcal{U}}^{n}_{1}}\times{{\mathcal{U}}^{n}_{2}}\times\cdots\times{{\mathcal{U}}^{n}_{K}}\| + {1 \over n}
\triangleq \de_n, \label{Fano_inequality}
\end{align}
and $\de_n \rightarrow 0$, where convergence is uniform in $d_{1}^{K+1}$ by \eqref{main_error_probability}.

Now, we can upper bound $H(U_{\mS}\vert U_{\mSc})$ as follows:
\begin{align}
H(U_{\mS}\vert U_{\mSc}) & = {1 \over n} H(\Us \vert \Usc, d_{1}^{K+1}) \nonumber \\
& \stackrel{(\rm a)}{=} {1 \over n} H(\Us \vert \Usc, \X{\mathcal{S}^{c}}{n}, d_{1}^{K+1}) \nonumber \\
& = {1 \over n} I(\Us; Y_{\mathsf{D}}^{n+\dm} \vert \Usc, \X{\mathcal{S}^{c}}{n}, d_{1}^{K+1}) + {1 \over n} H(\Us \vert Y_{\mathsf{D}}^{n+\dm}, \Usc, \X{\mathcal{S}^{c}}{n}, d_{1}^{K+1}) \nonumber \\
& \stackrel{(\rm b)}{\leq} {1 \over n} I(\X{\mS}{n}; Y_{\mathsf{D}}^{n+\dm} \vert \Usc, \X{\mathcal{S}^{c}}{n}, d_{1}^{K+1}) + \de_{n} \nonumber \\ \nonumber
& \stackrel{(\rm c)}{=} {1 \over n} h(Y_{\mathsf{D}}^{n+\dm} \vert \Usc, \X{\mathcal{S}^{c}}{n}, d_{1}^{K+1}) - {1 \over n} h(Y_{\mathsf{D}}^{n+\dm} \vert \Usc,X^{n}_{[1,K+1]}, d_{1}^{K+1}) + \de_{n} \nonumber \\
& \stackrel{(\rm d)}{\leq} {1 \over n} h(Y_{\mathsf{D}}^{n+\dm} \vert \X{\mathcal{S}^{c}}{n}, d_{1}^{K+1}) - {1 \over n} h(Y_{\mathsf{D}}^{n+\dm} \vert \Usc,X_{[1,K+1]}^{n}, d_{1}^{K+1}) + \de_{n} \nonumber \\ \nonumber
&={1\over n} h(\big\{\sum_{\ell=1}^{K+1}g_{\ell\mathsf{D}}X_{\ell}[i-d_{\ell}]+Z_{\mathsf{D}}[i]\big\}_{i=0}^{n+\dm-1}\vert \X{\mathcal{S}^{c}}{n}, d_{1}^{K+1})-{1\over n}h(Z_{\mathsf{D}}^{n+\dm})+\delta_{n}\\ \nonumber
&={1\over n} h(\big\{\sum_{\ell\in \mathcal{S}}g_{\ell\mathsf{D}}X_{\ell}[i-d_{\ell}]+Z_{\mathsf{D}}[i]\big\}_{i=0}^{n+\dm-1}\vert \X{\mathcal{S}^{c}}{n}, d_{1}^{K+1})-{1\over n}h(Z_{\mathsf{D}}^{n+\dm})+\delta_{n} \\
& \leq {1 \over n} h(Y_{\mathsf{D}(\mS)}^{n+\dm} \vert d_{1}^{K+1}) - {1 \over n}h(Z_{\mathsf{D}}^{n+\dm}) + \de_{n} \nonumber \\
& = {1 \over n} I(\X{\mS}{n}; Y_{\mathsf{D}(\mS)}^{n+\dm} \vert d_{1}^{K+1}) + \de_{n} \label{new_MARC_equation}
\end{align}
\noindent where in $(\rm a)$ we used the fact that $\X{\mSc}{n}$ is a function of only ${U}_{\mSc}^{n}$, in $(\rm b)$ we used the data processing inequality and \eqref{Fano_inequality}, in $(\rm c)$ we used $X^{n}_{[1,K+1]}$ based on the definition in \eqref{Eq:definitionofXs}, and lastly in $(\rm d)$ we made use of the fact that conditioning does not increase the entropy.

But \eqref{new_MARC_equation} represents the mutual information at the destination's output of the Gaussian sliced TA-MARC $\mathcal{M}(\mS)$ corresponding to the original Gaussian TA-MARC. Thus, using Lemma \ref{Key_lemma}, we can now further upper bound the mutual information term in \eqref{new_MARC_equation} by the corresponding mutual information term in the corresponding sliced cyclic MARC and derive
\begin{align}
H(U_{\mS}\vert U_{\mSc}) \leq  {1 \over n} I(\X{\mS}{n}; {\tilde{Y}_{\mathsf{D}(\mS)}}^{n} \vert d_{1}^{K+1}) + \ep_{n} + \de_{n}. \label{after_lemma}
\end{align}

Now, let $D_{\ell}, \ell=1,\cdots,K+1,$ be a sequence of independent random variables that are each uniformly distributed on the set $\{0,1,\cdots,\dm(n)\}$ and also independent of $\{U^{n}_{\ell}\}_{\ell=1}^{K+1}$, $\{Z_{\mathsf{D}}[i]\}_{i=0}^{n-1}$, and $\{Z_{\mathsf{R}}[i]\}_{i=0}^{n-1}$. Since
\eqref{after_lemma} is true for every choice of $d_{1}^{K+1} \in \{0,1,\cdots,\dm(n)\}^{K+1}$, $H(U_{\mS}\vert U_{\mSc})$ can
also be upper bounded by the average over $d_{1}^{K+1}$ of $I(\X{\mS}{n}; {\tilde{Y}_{\mathsf{D}(\mS)}}^{n} \vert d_{1}^{K+1})$. Hence,
\begin{align}
H(U_{\mS}\vert U_{\mSc}) & \leq I(\X{\mS}{n}; {\tilde{Y}_{\mathsf{D}(\mS)}}^{n} \vert D_{1}^{K+1})+\epsilon_{n}+\delta_{n} \nonumber \\
& \stackrel{\rm(a)}{=}  I(\X{\mS}{n}; \hat{\tilde{Y}}_{\mathsf{D}(\mS)}^{n} \vert D_{1}^{K+1}) +\epsilon_{n}+\delta_{n}, \label{eq:middle_step}
\end{align}
\noindent where $\hat{\tilde{Y}}_{\mathsf{D}(\mS)}^{n} = {\rm{DFT}}({\tilde{Y}}_{\mathsf{D}(\mS)}^{n})$, and $\rm(a)$ follows from the
fact that the DFT is a bijection.

Expanding $I(\X{\mS}{n}; \hat{\tilde{Y}}_{\mathsf{D}(\mS)}^{n} \vert D_{1}^{K+1})$ in the right hand side of \eqref{eq:middle_step},
\begin{align}
H(U_{\mS}\vert U_{\mSc}) & \leq {1 \over n} [h(\hat{\tilde{Y}}_{\mathsf{D}(\mS)}^{n} \vert D_{1}^{K+1}) - h(\hat{\tilde{Y}}_{\mathsf{D}(\mathcal{S})}^{n} \vert X_{\mS}^{n}, D_{1}^{K+1})] + \ep_{n} + \de_{n} \nonumber \\
& \leq {1 \over n} [h(\hat{\tilde{Y}}_{\mathsf{D}(\mS)}^{n}) - h(\hat{Z}_{\mathsf{D}}^{n})]  + \ep_{n} + \de_{n}, \nonumber
\end{align}
\noindent where $\hat{Z}_{\mathsf{D}}^{n}={\rm{DFT}}(Z_{\mathsf{D}}^{n})$ has {i.i.d.} entries with $\hat{Z}_{\mathsf{D}}[i] \sim
\mathcal{C}\mathcal{N}(0,N)$. Recall $\hat{X}_{{\ell}}^{n} = {\rm{DFT}}(X_{{\ell}}^{n})$. Then,
\begin{align}
h(\hat{\tilde{Y}}_{\mathsf{D}(\mS)}^{n}) & = h\left(\sum_{\ell \in \mS} {e^{-j\Bt(D_{{\ell}})}} \odot g_{{\ell}\mathsf{D}}\hat{X}_{{\ell}}^{n} + \hat{Z}_{\mathsf{D}}^{n} \right) \nonumber \\
& \leq \sum_{i=0}^{n-1} h\left(\sum_{\ell \in \mS} {e^{-j2\pi i {D_{{\ell}}}\over n}} g_{{\ell}\mathsf{D}} \hat{X}_{{\ell}}[i] + \hat{Z}_{\mathsf{D}}[i]\right), \nonumber
\end{align}
\noindent where ${e^{-j\Bt(D)}} \triangleq (e^{-j2\pi i D \over n})_{i=0}^{n-1}$ is an $n$-length vector, and $\odot$ denotes
element-wise vector multiplication. Thus,
\begin{align}
H(U_{\mS}\vert U_{\mSc})
&\leq {1 \over n} \sum_{i=0}^{n-1} \left[h\left(\sum_{\ell \in \mS} {e^{-j2\pi i {D_{{\ell}}}\over n}} g_{\ell\mathsf{D}} \hat{X}_{{\ell}}[i] + \hat{Z}_{\mathsf{D}}[i]\right)- h(\hat{Z}_{\mathsf{D}}[i])\right] + \ep_{n} + \de_{n} \nonumber\\
& \leq {1 \over n} \sum_{i=0}^{n-1} \log\left(1 + { { {\mathbb{E}\left\vert \sum_{\ell \in \mS} {e^{-j2\pi i {D_{{\ell}}}\over n}} g_{\ell\mathsf{D}} \hat{X}_{{{\ell}}}[i] \right\vert^{2}}} \over {N}}\right)
+ \ep_{n} + \de_{n}. \label{before_split_into_three}
\end{align}

We now divide the sum in \eqref{before_split_into_three} into three terms for $0 \leq i \leq \al(n)-1$, $\al(n) \leq i \leq n-\al(n)-1$, and $n-\al(n) \leq i \leq n-1$, where $\al(n): \mathbb{N} \rightarrow \mathbb{N}$ is a function such that
\begin{align}
{\al(n) \over n} \rightarrow 0, \ \ {\al(n)\dm(n) \over n} \rightarrow \infty. \label{condition_on_alpha}
\end{align}
\noindent An example of such an $\al(n)$ is the function $\alpha(n) = \lceil {n \over
\dm(n)}{\log\dm(n)} \rceil $. Consequently, we first upper bound the tail terms and afterwards the main term in the sequel.

For the terms in $0\leq i \leq \al(n)-1$, we have

\begin{align}
\nonumber
{1 \over n} \sum_{i=0}^{\al(n)-1} \log\left(1 + { { {\mathbb{E}\left\vert \sum_{\ell \in \mS} {e^{-j2\pi i {D_{{\ell}}}\over n}} g_{\ell\mathsf{D}} \hat{X}_{{{\ell}}}[i] \right\vert^{2}}} \over {N}}\right) & \stackrel{\rm(a)}{\leq}  {1 \over n} \sum_{i=0}^{\al(n)-1} \log\left(1 + { {\sum_{\ell\in \mS}\vert g_{\ell\mathsf{D}}\vert^{2} \cdot \sum_{\ell \in \mS}\mathbb{E} \vert \hat{X}_{\ell}[i]\vert^{2} } \over {N}}\right) \\ \nonumber
& \stackrel{\rm(b)}{\leq} {\al(n) \over n}  \log\left(1 + { {\sum_{i=0}^{\al(n)-1} \left[\sum_{\ell\in \mS}\vert g_{\ell\mathsf{D}}\vert^{2}\cdot \sum_{\ell \in \mS}\mathbb{E} \vert \hat{X}_{\ell}[i]\vert^{2} \right]} \over {\alpha(n)N}}\right)\\ \nonumber
\nonumber
& \stackrel{\rm(c)}{\leq}{\al(n) \over n} \log\left(1 +{n\over \alpha(n)} {{ \sum_{\ell\in \mS}\vert g_{\ell\mathsf{D}}\vert^{2} \cdot \sum_{\ell \in \mS}P_{\ell}} \over {N}}\right) \\
& \triangleq \lambda_{n}, \label{1st_sum}
\end{align}
\noindent where $(\rm a)$ follows by the Cauchy-Schwartz inequality (cf. \eqref{eq_Cauchy}), $(\rm b)$ follows by the concavity of the $\log$ function and $(\rm c)$ follows by the power constraints \eqref{Power_constraint}.
Also, for $n-\al(n) \leq i \leq n-1$, a similar upper bound can be derived by the symmetry of the problem as follows

\begin{align}
{1 \over n} \sum_{i=n-\al(n)}^{n-1} \log\left(1 + { { {\mathbb{E}\left\vert \sum_{\ell \in \mS} {e^{-j2\pi i {D_{{\ell}}}\over n}} g_{\ell\mathsf{D}} \hat{X}_{{{\ell}}}[i] \right\vert^{2}}} \over {N}}\right) \leq \lambda_{n}. \label{2nd_sum}
\end{align}

To bound the third component of \eqref{before_split_into_three} for $\alpha(n)\leq i\leq n-\alpha(n)-1$, we first obtain that
\begin{align}
{ {\mathbb{E}\left\vert \sum_{\ell \in \mS} {e^{-j2\pi i {D_{{\ell}}}\over n}} g_{\ell\mathsf{D}} \hat{X}_{{{\ell}}}[i] \right\vert^{2}}}
 = \sum_{\ell \in \mS} \vert g_{\ell\mathsf{D}}\vert^{2} \mathbb{E}\vert \hat{X}_{\ell}[i] \vert^{2} + {\sum_{ \substack {(\ell,\ell^{'}) \in \mS^{2} \\ \label{Eq:dependent_terms} \ell < \ell^{'}} }} 2\Re\mathbb{E}\left\{{e^{-j2\pi i ({D_{{\ell}} - D_{{\ell^{'}}}})\over n}} g_{\ell\mathsf{D}}g^{*}_{\ell'\mathsf{D}} \hat{X}_{{{\ell}}}[i]\hat{X}^{*}_{{{\ell^{'}}}}[i] \right\},
\end{align}
\noindent where $\Re(z)$ is the real part of $z\in \mathbb{C}$. Now, the following two cases can occur

$i$) $\ell< \ell'<K+1$: In this case, both $\hat{X}_{\ell}[i]$ and $\hat{X}^{*}_{\ell'}[i]$ are independent of $D_{\ell}$ and $D_{\ell'}$.

$ii$) $\ell <\ell'=K+1$: In this case, $\hat{X}_{\ell}[i]$ and $\hat{X}^{*}_{\ell'}[i]$ are independent of $D_{\ell'}$ . However, $\hat{X}^{*}_{\ell'}[i]$, that corresponds to the channel input of the relay, is a function of $\{Y_{\mathsf{R}}[0],Y_{\mathsf{R}}[1],\cdots,Y_{\mathsf{R}}[i-1]\}$ and is thus correlated with delays of all source node transmitters, \textit{i.e.}, $D_{\ell},\ell=1,2,\cdots, K$, due to \eqref{channel-model-2}.

In either scenario, we can proceed from \eqref{Eq:dependent_terms} by separating $e^{j2\pi iD_{\ell'}\over n}$ from the remaining terms inside the expectation. Specifically,
\begin{align}
\nonumber
\mathbb{E}\left\vert \sum_{\ell \in \mS} {e^{-j2\pi i {D_{{\ell}}}\over n}} g_{\ell\mathsf{D}} \hat{X}_{{{\ell}}}[i] \right\vert^{2} &=
\sum_{\ell \in \mS} \vert g_{\ell\mathsf{D}}\vert^{2} \mathbb{E}\vert \hat{X}_{\ell}[i] \vert^{2} + {\sum_{ \substack {(\ell,\ell^{'}) \in \mS^{2} \\  \ell < \ell^{'}} }} 2\Re\left(\mathbb{E} \left\{ {e^{j2\pi i {D_{{\ell'}}}\over n}}\right\}\mathbb{E}\left\{e^{-j2\pi i {D_{{\ell}}}\over n} g_{\ell\mathsf{D}}g^{*}_{\ell'\mathsf{D}} \hat{X}_{{{\ell}}}[i]\hat{X}^{*}_{{{\ell^{'}}}}[i] \right\}\right)
\\ \nonumber
&\hspace{-15mm}\leq \sum_{\ell \in \mS} \vert g_{\ell\mathsf{D}}\vert^{2} \mathbb{E}\vert \hat{X}_{\ell}[i] \vert^{2} + {\sum_{ \substack {(\ell,\ell^{'}) \in \mS^{2} \\  \ell < \ell^{'}} }} 2\left\vert\mathbb{E} \left\{ {e^{j2\pi i {D_{{\ell'}}}\over n}}\right\}\mathbb{E}\left\{e^{-j2\pi i {D_{{\ell}}}\over n} g_{\ell\mathsf{D}}g^{*}_{\ell'\mathsf{D}} \hat{X}_{{{\ell}}}[i]\hat{X}^{*}_{{{\ell^{'}}}}[i] \right\}\right\vert \\ \nonumber
&\hspace{-15mm} = \sum_{\ell \in \mS} \vert g_{\ell\mathsf{D}}\vert^{2} \mathbb{E}\vert \hat{X}_{\ell}[i] \vert^{2} + \sum_{ \substack {(\ell,\ell^{'}) \in \mS^{2} \\ \ell < \ell^{'}} } 2 \vert g_{\ell\mathsf{D}}\vert \vert g_{\ell'\mathsf{D}}\vert \Bigg\vert\mathbb{E} \left\{ {e^{j2\pi i {D_{{\ell'}}}\over n}}\right\}\Bigg\vert\Bigg\vert\mathbb{E}\left\{{e^{-j2\pi i {D_{{\ell}}}\over n}} \hat{X}_{{{\ell}}}[i]\hat{X}^{*}_{{{\ell^{'}}}}[i]\right\}\Bigg\vert\\
&\hspace{-15mm} \stackrel{\rm (a)}{\leq}\sum_{\ell \in \mS} \vert g_{\ell\mathsf{D}}\vert^{2} \mathbb{E}\vert \hat{X}_{\ell}[i] \vert^{2}+ {1 \over \dm(n){\vert \sin({\pi i \over n}) \vert}} \sum_{ \substack {(\ell,\ell^{'}) \in \mS^{2} \\ \nonumber \ell < \ell^{'}} } \vert g_{\ell\mathsf{D}}\vert \vert g_{\ell'\mathsf{D}}\vert  {\left(\mathbb{E}\vert \hat{X}_{\ell}[i] \vert^{2} + \mathbb{E}\vert \hat{X}_{\ell'}[i] \vert^{2} \right)} \\
&\hspace{-15mm} \stackrel{\rm (b)}{\leq}\sum_{\ell \in \mS} \vert g_{\ell\mathsf{D}}\vert^{2} \mathbb{E}\vert \hat{X}_{\ell}[i] \vert^{2}+ {1 \over \dm(n){\vert \sin({\pi \alpha(n) \over n}) \vert}} \sum_{ \substack {(\ell,\ell^{'}) \in \mS^{2} \\  \ell < \ell^{'}} } \vert g_{\ell\mathsf{D}}\vert \vert g_{\ell'\mathsf{D}}\vert  {\left(\mathbb{E}\vert \hat{X}_{\ell}[i] \vert^{2} + \mathbb{E}\vert \hat{X}_{\ell'}[i] \vert^{2} \right)},
\label{received_power_ultimate_upperbound}
\end{align}
\noindent where the derivation of $\rm (a)$ is presented in Appendix \ref{proof of correlation_inequality}, and $\rm (b)$ follows from the inequality
\begin{align}
\sin({\pi \alpha(n)\over n})\leq \sin({\pi i\over n}),\hspace{2mm} \text{for all} \ i\in [\alpha(n),n-\alpha(n)-1].
\end{align}
By summing \eqref{received_power_ultimate_upperbound} over $\al(n) \leq i \leq n - \al(n)-1$, we further obtain
\begin{align}
\nonumber
&\sum_{i=\al(n)}^{n-\al(n)-1}{{\mathbb{E}\left\vert \sum_{\ell \in \mS} {e^{-j2\pi i {D_{{\ell}}}\over n}} g_{\ell\mathsf{D}} \hat{X}_{{{\ell}}}[i] \right\vert^{2}}}\\
\nonumber &\quad \leq \sum_{i=\al(n)}^{n-\al(n)-1}\sum_{\ell \in \mS} \vert g_{\ell\mathsf{D}}\vert^{2} \mathbb{E}\vert \hat{X}_{\ell}[i] \vert^{2}+ {1 \over \dm(n){\vert \sin({\pi \alpha(n) \over n}) \vert}} \sum_{i=\al(n)}^{n-\al(n)-1}\sum_{ \substack {(\ell,\ell^{'}) \in \mS^{2} \\  \ell < \ell^{'}} } \vert g_{\ell\mathsf{D}}\vert \vert g_{\ell'\mathsf{D}}\vert  {\left(\mathbb{E}\vert \hat{X}_{\ell}[i] \vert^{2} + \mathbb{E}\vert \hat{X}_{\ell'}[i] \vert^{2} \right)}\\ \nonumber
&\quad \stackrel{\rm (a)}{\leq} \sum_{\ell \in \mS} \vert g_{\ell\mathsf{D}}\vert^{2} nP_{\ell}+ {1 \over \dm(n){\vert \sin({\pi \alpha(n) \over n}) \vert}} \sum_{ \substack {(\ell,\ell^{'}) \in \mS^{2} \\  \ell < \ell^{'}} } \vert g_{\ell\mathsf{D}}\vert \vert g_{\ell'\mathsf{D}}\vert (nP_{\ell}+nP_{\ell'})\\
& \quad = n\left[\sum_{\ell \in \mS} \vert g_{\ell\mathsf{D}}\vert^{2} P_{\ell}+ {\zeta(\mS) \over \dm(n){\vert \sin({\pi \alpha(n) \over n}) \vert}} \right],\label{Eq:wierd}
\end{align}
where $\rm (a)$ is due to the power constraint in \eqref{Power_constraint}, and
\begin{align}
\zeta(\mS)\triangleq  \sum_{ \substack {(\ell,\ell^{'}) \in \mS^{2} \\  \ell < \ell^{'}} } \vert g_{\ell\mathsf{D}}\vert \vert g_{\ell'\mathsf{D}}\vert (P_{\ell}+P_{\ell'}).
\end{align}
Based on the result in \eqref{Eq:wierd}, we upper bound the third component of \eqref{before_split_into_three} as below
\begin{align}
\nonumber
&{1 \over n} \sum_{i=\al(n)}^{n - \al(n)-1} \log\left(1 + {{ {\mathbb{E}\left \vert \sum_{\ell \in \mS} {e^{-j2\pi i {D_{{\ell}}}\over n}} g_{\ell\mathsf{D}} \hat{X}_{{{\ell}}}[i] \right\vert^{2}} } \over {N}}\right) \\
\nonumber
& \quad \stackrel{\rm(a)} \leq {n-2\alpha(n) \over n} \log\left(1 +  {\sum_{i=\al(n)}^{n-\al(n)-1} \left[{\mathbb{E}\left \vert \sum_{\ell \in \mS} {e^{-j2\pi i {D_{{\ell}}}\over n}} g_{\ell\mathsf{D}} \hat{X}_{{{\ell}}}[i] \right\vert^{2}}\right]  \over {N}(n-2\alpha(n))} \right)
\\  \label{3rd_sum}
& \quad \stackrel{\rm(b)}\leq {n-2\alpha(n) \over n} \log\left(1 +{n\over n-2\alpha(n)}  {\sum_{\ell \in \mS} \vert g_{\ell\mathsf{D}}\vert^{2} P_{\ell}+ {\zeta (\mS) \over \dm(n){\vert \sin({\pi \alpha(n) \over n}) \vert}} \over {N}} \right),
\end{align}
\noindent where $\rm(a)$ follows by the concavity of the $\log$ function, and $\rm(b)$ follows from \eqref{Eq:wierd}.

Now, by combining \eqref{before_split_into_three}, \eqref{1st_sum}, \eqref{2nd_sum}, and \eqref{3rd_sum} we derive
\begin{align}
\label{Eq:beforetakinglimit}
H(U_{\mS}\vert U_{\mSc})
&\leq {n-2\alpha(n) \over n}  \log\left(1 +{n\over n-2\alpha(n)}  {\sum_{\ell \in \mS} \vert g_{\ell\mathsf{D}}\vert^{2} P_{\ell}+ {\zeta (\mS) \over \dm(n){\vert \sin({\pi \alpha(n) \over n}) \vert}} \over {N}}\right) +
2\lambda_{n} + \ep_{n} + \de_{n}.
\end{align}

To obtain the asymptotic bound, we recall that due to the choice of $\al(n)$ in \eqref{condition_on_alpha},
\begin{align}
{n-2 \alpha(n)\over n}&\rightarrow 1, \nonumber \\
\sin\left({{\pi \alpha(n)}\over{n}}\right)/{\pi \alpha(n)\over n}& \rightarrow 1, \nonumber\\
{{1} \over {\dm(n){\vert \sin({{{\pi}\al(n)}\over n}) \vert}}} & \rightarrow {{n \over \pi\dm(n)\al(n)}} \rightarrow 0, \nonumber
\end{align}
\noindent as $n\rightarrow \infty$. Therefore, it can be easily verified from \eqref{Eq:beforetakinglimit} that since $\zeta(\mS)<\infty$, and $\lambda_{n},\delta_{n},\epsilon_{n}\rightarrow 0$ as $n\rightarrow \infty$,
\begin{align}
H(U_{\mathcal{S}}\vert U_{\mathcal{S}^{c}})\leq \log\left(1 +{{\sum_{\ell \in \mS} \vert g_{\ell\mathsf{D}}\vert^{2} P_{\ell}} \over {N}}\right),
\end{align}
where we recall that the subset $\mS\subseteq [1,K+1]$ includes the relay, {i.e.}, $\{K+1\}\in \mS$.
\end{IEEEproof}

\vspace{-.3cm}
\section{Achievability}\label{section_achievability}

We now focus on demonstrating the sufficiency of the condition that was proved to be a necessary condition for reliable communication in Lemma \ref{Lemma:Reliable_Communication} and thus conclude that the region described by \eqref{separation_TAMAC_1} is indeed the
JSCC capacity region. To establish the achievability argument, we follow a {\em tandem} (separate)
source-channel coding scheme. Thus, the communication process will be divided into two parts:
source coding and channel coding. In the sequel, we simply state the results for each of both
source and channel coding, and finally by combining them prove the achievability lemma.

{\em Source Coding}: From Slepian-Wolf coding \cite{Slepian_Wolf_main:1973}, for the correlated
source $(U_{1}^{n},U_{2}^{n},\cdots,U_{K}^{n})$, if we have $K$ $n$-length sequences of source codes with rates $(R_1,R_2,\cdots,R_{K})$, for
asymptotically lossless representation of the source, we should have
\begin{align}\label{Source_coding1}
H(U_{\mS}\vert U_{\mSc}) &< R_{\mS}, \quad \forall \mathcal{S} \subseteq [1,K+1]:\{K+1\} \in \mS ,
\end{align}
\noindent where by definition $R_{\mS} \triangleq \sum_{\ell \in \mathcal{S}} R_\ell$, $R_{K+1} \triangleq 0$, and $U_{K+1} \triangleq \emptyset$.

{\em Channel Coding}: Next, for fixed source codes with rates $(R_1,R_2,\cdots,R_K)$, we make channel codes
for the TA-MARC separately such that the channel codes can be reliably decoded at the receiver side.
In particular, we use the block Markov coding scheme used in \cite{Kramer_MARC_Main1:2004} on top of the coding strategy used in \cite{Cover_McEliece:81}, in order to make reliable channel codes. Indeed, we directly apply the decoding technique of \cite{Cover_McEliece:81} to a series of block Markov codes which results in an achievable rate region equivalent to the intersection of two MACs with encoders of the transmitters with indices $1,\cdots,K$, and all transmitters, and decoders of the relay and destination respectively. In the sequel, we briefly give some details of the block Markov coding scheme and the coding strategy for the delayed codewords.

\begin{table}
\centering
\begin{tabular}{|c|c|c|c|c|c|}
  \hline
 Encoder  & Block $1$ & Block $2$ & $\cdots$ &  Block $B$ & Block $B+1$\\
  \hline       \hline
    $1$ & $\bx{1}(1,W_{11})$ & $\bx{1}(W_{11},W_{12})$ & $\cdots$ & $\bx{1}(W_{1(B-1)},W_{1B})$
&  $\bx{1}(W_{1B},1)$ \\
  \hline       \hline
  $\vdots$ & $\vdots$ & $\vdots$ & $\cdots$ &$\vdots$ & $\vdots$ \\
  \hline       \hline
    $K$ & $\bx{K}(1,W_{K1})$ & $\bx{K}(W_{K1},W_{K2})$ & $\cdots$ &$\bx{K}(W_{K(B-1)},W_{KB})$ & $\bx{K}(W_{KB},1)$ \\
 \hline          \hline
$K+1$ & $\bx{K+1}(1,\cdots,1)$ & $\bx{K+1}(W_{11},\cdots,W_{K1})$ & $\cdots$ & $\bx{K+1}(W_{1(B-1)},\cdots,W_{K(B-1)})$ &$\bx{K+1}(W_{1B},\cdots,W_{KB})$\\
 \hline
\end{tabular}
\caption{Block Markov encoding scheme for the Gaussian TA-MARC.}\label{Table_TAMARC}
\end{table}

\begin{itemize}
\item {\em Block Markov coding}: Table I shows the block Markov coding configuration used to transmit the codewords of the encoders of the Gaussian TA-MARC. First fix a distribution $p(x_{1})\cdots p(x_{K+1})$ and construct random codewords $\bx{1},\cdots, \bx{K+1}$ based on the corresponding distributions. The message $W_{i}$ of each encoder is divided to $B$ blocks $W_{i1},W_{i2},\cdots,W_{iB}$ of
$2^{nR_{i}}$ bits each, $i = 1,\cdots,K$. The codewords are transmitted in $B + 1$ blocks based on the block Markov
encoding scheme depicted in Table I. After each block, the relay makes a MAC decoding and uses the decoded messages $W_{1(i-1)},\cdots,W_{K(i-1)}$ to send the codewords in the next block. Also, the decoding at the destination is performed at the end of the last block and in a backward block-by-block manner, also known as {\em backward decoding} \cite{Kramer_MARC_Main1:2004}. We let $B \rightarrow \infty$ to approach the original rates $R_1,\cdots,R_K$.

\item {\em Coding strategy of \cite{Cover_McEliece:81}}: The encoders transmit their codewords as shown in Table I and in $B$ blocks, albeit with delays $d_{1},\cdots,d_{K+1}$. Note that if the MARC was synchronous, one would obtain the achievable rate region resulting from the intersection of two MACs. However, using a simply generalized version of the coding strategy used in \cite{Cover_McEliece:81}, it can be seen that the same region is achievable for the time asynchronous case. In particular, at the end of the $i$th block, the relay decoder inspects the received vector $Y^{n+\dm(n)}_{\sf{R}}$ for the presence of codewords $\bx{1}(W_{1i}),\cdots,\bx{K}(W_{Ki})$, embedded in it with arbitrarily shifts. Likewise, at the end of the last block, the destination decoder inspects the received vector $Y^{n+\dm(n)}_{\sf{D}}$ to first decode $W_{1B},\cdots,W_{KB}$ and consequently decode the previous messages in a backward manner. In all of these decoding cases, like \cite{Cover_McEliece:81}, we look for the codewords under all possible shifts up to the maximum delay $\dm$ such that the shifted codewords and the $(n+\dm)$-length received vector are jointly typical. Therefore, the decoders at the relay and destination need to look for $\dm(n)^{K}$, and $\dm(n)^{K+1}$ combination of codewords respectively and find the one that is jointly typical with $Y^{n+\dm(n)}_{\sf{R}}$ or $Y^{n+\dm(n)}_{\sf{D}}$. Following similar error analysis as in \cite{Cover_McEliece:81}, now for a $K$ user system with $K$ delays, and due to the assumption that $\dm(n) / n \rightarrow 0$, it can be seen that the standard synchronous $K$ user MAC capacity constraints are derived in order to achieve asymptotically vanishing probability of error.
\end{itemize}

Hence, for reliable communication of the source indices over the Gaussian TA-MARC, the following sets of inequalities that represents MAC decoding at the relay and destination should then be satisfied:
\begin{align}
R_{\mS} < I(X_{\mS};Y_{\sf{R}} \vert X_{\mS^{c}}), \quad \forall \mS \subseteq [1,K], \label{eq:decoding_relay}
\end{align}
\noindent and
\begin{align}
R_{\mS} < I(X_{\mS};Y_{\sf{D}} \vert X_{\mS^{c}}), \quad \forall \mS \subseteq [1,K+1]: \{K+1\} \in \mS, \label{eq:decoding_destination}
\end{align}
\noindent for an input distribution $p(x_{1})\cdots p(x_{K+1})$.

By choosing Gaussian input distributions, the constraints in \eqref{eq:decoding_relay}-\eqref{eq:decoding_destination} will be reduced to logarithmic rate functions. It is then straight forward to see that under the gain conditions
\begin{align}\label{gain_conditions}
\sum_{\ell \in \mathcal{S}} \vert g_{\ell \mathsf{R}}\vert^{2}P_{\ell} \geq {\vert g_{(K+1){\sf{D}}}\vert}^{2} P_{K+1}+ \sum_{\ell \in \mS} \vert g_{{\ell}\mathsf{D}}\vert ^{2}P_{{\ell}}, \quad \forall \mathcal{S} \subseteq [1,K],
\end{align}
\noindent the destination decoding constraints \eqref{eq:decoding_destination} will dominate \eqref{eq:decoding_relay}, and we can thus derive the following conditions on $R_1,\cdots,R_K$, as sufficient conditions for reliable communication of source coded indices over a Gaussian TA-MARC:
\begin{align}
\sum_{\ell \in \mathcal{S}} R_\ell & < \log\left(1+{\sum_{\ell \in \mS }\vert g_{\ell \sf{D}}\vert ^{2}P_{\ell} \over N}\right), \quad \forall \mathcal{S} \subseteq [1,K+1]: \{K+1\} \in \mS. \label{Channel_coding1}
\end{align}

\begin{lemma}
A sufficient condition for reliable communication of the source $(U^{n}_1,\cdots,U^{n}_K)$ over the TA-MARC defined by \eqref{channel-model-1}-\eqref{channel-model-2}, and under gain conditions of \eqref{gain_conditions}, is given by \eqref{separation_TAMAC_1}, with $\leq$ replaced by $<$.
\end{lemma}

\begin{IEEEproof}

From \eqref{separation_TAMAC_1}, it can be seen that there exist
choices of $R_1,\cdots,R_2$ such that the Slepian-Wolf conditions
\eqref{Source_coding1} and the channel coding conditions
\eqref{Channel_coding1} are simultaneously satisfied. Since error
probabilities of both the source coding part and channel coding part vanish asymptotically, then
the error probability of the combined tandem scheme also vanishes asymptotically and the proof of
the lemma is complete.
\end{IEEEproof}
%
%
%
%
%
\vspace{-.2cm}
\section{Separation Theorems}\label{results_statements}

Based on the converse and achievabaility results presented in Sections \ref{section_converse} and \ref{section_achievability}, we can now combine the results and state the following separation theorem for a Gaussian TA-MARC

\begin{theorem}\label{main_theorem}
{\em Reliable Communication over a Gaussian TA-MARC}: Consider a Gaussian TA-MARC with the gain conditions \eqref{gain_conditions}. Then, necessary conditions for reliably sending a source $(U_{1}^{n},\cdots,U_{K}^{n}) \sim {\prod_{i}}p(u_{1i},\cdots,u_{Ki})$, over such a TA-MARC are given by \eqref{separation_TAMAC_1}. Furthermore, \eqref{separation_TAMAC_1}, with $\leq$ replaced by $<$, also gives a sufficient condition for reliable communications over such a TA-MARC and can be achieved by separate source-channel coding. \thmend 
\end{theorem}

\begin{figure}
\centering{\includegraphics[keepaspectratio = true, width=10.5cm]{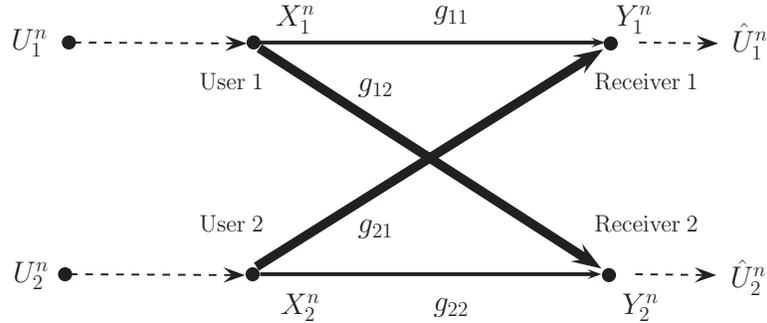}}
\caption{Gaussian Time-Asynchronous Interference Channel (TA-IC) with Strong Interference Gains.}
\label{fig:TA-IC}
\end{figure}

Theorem \ref{main_theorem} can be easily specialized to a MAC if we impose $P_{K+1} = 0$ and eliminate the role of the relay. Thus, the result of \cite{Saffar_Mitran_ISIT2013} for a $2$-user TA-MAC is a direct consequence of Theorem \ref{main_theorem}. As a result, we can also state the following corollary for a Gaussian time asynchronous interference channel (TA-IC) with strong interference conditions depicted in Fig. \ref{fig:TA-IC}.  The result of the corollary is based on the fact that in the strong interference regime, the Gaussian interference channel can be reduced to the intersection of two Gaussian MACs with no loss. Namely, if each receiver can correctly decode its own channel input sequence, in the strong interference regime, it can also correctly decode the other channel input sequence (see \cite{Sato:1981} for details). In the context of JSCC, we note that by using the strong interference conditions and the one-to-one mappings between source and channel sequences, one can argue that both of the receivers can recover both source sequences $U^{n}_{1}$, $U^{n}_{2}$ provided there are encoders and decoders such that each receiver can reliably decode its {\em own} source sequence. Specifically, in the converse part, the first receiver can decode $U_1^n$ by assumption and this in turn enables it to reconstruct the channel input $X_1^n$ from $U_1^n$. Then, similar to \cite{Sato:1981}, from $X_1^n$ and $Y_1^n$, the first receiver constructs $\tilde{Y}_2^n = g_{12} X_1^n + {g_{22}\over g_{21}}(Y_1^n - g_{11}X_1^n) = g_{12} X_1^n + g_{22} X_2^n + \tilde{Z}_2^n,$ where the noise power of each $\tilde{Z}_2[i]$ is less than that of $Z_2[i]$. Receiver 1 can then reconstruct $U_2^n$ from $\tilde{Y}_2^n$ using receiver 2's decoder. Similarly, receiver 2 can also recover $U_1^n$. 
{Therefore, under the strong interference regime, necessary (resp. sufficient) conditions for JSCC are described by the intersection of the necessary (resp. sufficient) conditions of two MACs.}



\begin{corollary}
Necessary conditions for reliably sending arbitrarily
correlated sources $(U_1,U_2)$ over a TA-IC with strong interference conditions $\vert g_{11}\vert  \leq \vert g_{12}\vert,\vert g_{22} \vert \leq \vert g_{21} \vert$ are given by
\begin{align}
\mkern-13mu H(U_1 \vert U_2) & \leq \log(1+\vert g_{11} \vert^{2} P_1/N), \label{separation_1_IC}\\
\mkern-13mu H(U_2 \vert U_1) & \leq \log(1+\vert g_{22} \vert^{2} P_2/N), \label{separation_2_IC}\\
\mkern-13mu H(U_1,U_2)  & \leq \log(1+(\vert g_{11}\vert^{2}P_1+\vert g_{21}\vert^2P_2)/N), \label{separation_3_IC}\\
\mkern-13mu H(U_1,U_2)  & \leq \log(1+(\vert g_{12}\vert^{2}P_1+\vert g_{22}\vert^2P_2)/N), \label{separation_4_IC}
\end{align}
\noindent where $g_{ij}, i,j \in \{1,2\}$ represents the complex gain from node $i$ to the receiver $j$ in a two user interference channel. The same conditions \eqref{separation_1_IC}-\eqref{separation_3_IC} with $\leq$ replaced
by $<$ describe sufficient conditions for reliable communication. \thmend
\end{corollary}

\vspace{-.25cm}
\section{Conclusion}\label{section_conclusion}
The problem of sending arbitrarily correlated sources over a time asynchronous multiple-access
relay channel with maximum offset between encoders $\dm(n) \rightarrow \infty$, as $n\rightarrow \infty$,
is considered. Necessary and sufficient conditions for reliable communication are presented under
the assumption of ${\dm(n) / n} \rightarrow 0$. Namely, a general outer bound on the capacity region is
first derived and then is shown to match the separate source-channel coding achievable region under specific gain conditions. Therefore,
under the gain conditions, separation is shown to be optimal and as a result, joint source-channel coding is not necessary
under time asynchronism with these gain conditions.

\appendices

%

\section{}
\label{proof of correlation_inequality}

Since $D_{\ell'}$ has a uniform distribution over $\{0,1,\cdots, \dm\}$ we have
\begin{align}
\left\vert\mathbb{E} \left\{ {e^{j2\pi i {D_{{\ell'}}}\over n}}\right\}\right\vert&=\left\vert\sum_{d=0}^{\dm}{1\over \dm+1}{e^{{j2\pi i d}\over n}}\right\vert\\
&=\left\vert {1\over \dm+1}{{e^{{j2\pi i (\dm+1)}\over n}}-1\over {e^{{j2\pi i}\over n}}-1} \right\vert \\
&= \left\vert{1\over \dm+1}{\sin({{\pi i(\dm+1)}\over n})\over {\sin ({{\pi i}\over n}})}\right\vert \\
&\leq {1\over \dm {\vert\sin ({{\pi i}\over n}})\vert}.
\end{align}
Thus, we obtain the following inequality
\begin{align}
\nonumber
&\hspace{-20mm}\sum_{ \substack {(\ell,\ell^{'}) \in \mS^{2} \\ \ell < \ell^{'}} } 2 \vert g_{\ell\mathsf{D}}\vert \vert g_{\ell'\mathsf{D}}\vert \Bigg\vert\mathbb{E} \left\{ {e^{j2\pi i {D_{{\ell'}}}\over n}}\right\}\Bigg\vert\Bigg\vert\mathbb{E}\left\{{e^{-j2\pi i {D_{{\ell}}}\over n}} \hat{X}_{{{\ell}}}[i]\hat{X}^{*}_{{{\ell^{'}}}}[i]\right\}\Bigg\vert \\
& \leq {1\over \dm {\vert\sin ({{\pi i}\over n}})\vert}\sum_{ \substack {(\ell,\ell^{'}) \in \mS^{2} \\ \ell < \ell^{'}} } 2 \vert g_{\ell\mathsf{D}}\vert \vert g_{\ell'\mathsf{D}}\vert \mathbb{E}\left\{ \Bigg\vert{e^{-j2\pi i {D_{{\ell}}}\over n}} \hat{X}_{{{\ell}}}[i]\hat{X}^{*}_{{{\ell^{'}}}}[i]\Bigg\vert\right\} \\
& = {1\over \dm {\vert\sin ({{\pi i}\over n}})\vert}\sum_{ \substack {(\ell,\ell^{'}) \in \mS^{2} \\ \ell < \ell^{'}} } 2 \vert g_{\ell\mathsf{D}}\vert \vert g_{\ell'\mathsf{D}}\vert \mathbb{E}\left\{ \big\vert\hat{X}_{{{\ell}}}[i]\big\vert\big\vert\hat{X}^{*}_{{{\ell^{'}}}}[i]\big\vert\right\}\\
&\stackrel{\rm(a)}{\leq} {1\over \dm {\vert\sin ({{\pi i}\over n}})\vert}\sum_{ \substack {(\ell,\ell^{'}) \in \mS^{2} \\ \ell < \ell^{'}} } \vert g_{\ell\mathsf{D}}\vert \vert g_{\ell'\mathsf{D}}\vert (\mathbb{E}\vert\hat{X}_{{{\ell}}}[i]\vert^{2}+\mathbb{E}\vert\hat{X}_{{{\ell^{'}}}}[i]\vert^{2}),
\end{align}
\noindent where $\rm(a)$ follows by the geometric inequality $2\sqrt{ab} \leq {a+b}$ with $a=\vert \hat{X}_{\ell}[i]\vert^{2}$ and $b=\vert \hat{X}_{\ell'}[i]\vert^{2}=\vert \hat{X}^{*}_{\ell'}[i]\vert^{2}$.

\bibliographystyle{ieeetr}
\bibliography{Hamidreza_bib2}

\end{document}